\documentclass[preprint,12pt]{elsarticle}

\usepackage{hyperref}
\usepackage{amssymb}
\usepackage{amsmath}
\usepackage{geometry}
\usepackage{graphicx}
\usepackage{cleveref}
\usepackage{subfigure}
\usepackage{bm}
\usepackage{longtable}
\usepackage{multirow}
\usepackage{float}
\usepackage{fullpage}
\usepackage{booktabs}
\usepackage[numbers]{natbib}
\usepackage{color}
\usepackage{xcolor}
\usepackage{diagbox}
\usepackage{longtable}
\usepackage[figuresright]{rotating}

\usepackage{lineno}

\journal{$-$}

\begin{document}

\begin{frontmatter}



\title{Initial release styles have limited effects on the hydrodynamic dynamics of a self-propelled fin in the unsteady wakes}


\author[labela]{Peng Han}

\author[labelb]{Dong Zhang}
\author[labela]{Jun-Duo Zhang}
\author[labela]{Wei-Xi Huang\corref{cor}}
\cortext[cor]{Corresponding author.}
\ead{hwx@mail.tsinghua.edu.cn}

\address[labela]{AML, Department of Engineering Mechanics, Tsinghua University, Beijing, 100084, P.R.China}

\address[labelb]{Qingdao Innovation and Development Base, Harbin Engineering University, Qingdao 266000, PR China}

\begin{abstract}
Living fish may suddenly encounter upstream obstacles, join the queue of the fish schooling, or detect upstream flow in advance, resulting in interactions with environmental vortices that can be abrupt or develop gradually from an initial state. The impact of initial conditions on fish swimming behavior in unsteady upstream vortices remains an open question. This study employs a self-propelled flexible fin model, the immersed boundary method, and direct simulation to analyze the hydrodynamics and locomotion of fish swimming behind a bluff cylinder and within the schooling, under different initial gaps and release styles. Additionally, the above tests were conducted with both quiescent flow fields and fully developed unsteady flows as initial conditions. The results indicate that schooling self-propelled fins are more resilient to initial perturbations compared to fins swimming behind a bluff body. More importantly, when simulations begin with a fully developed wake pattern, which better reflects natural environments, the characteristics of the self-propelled fins remain consistent regardless of the initial release styles. Therefore, from a hydrodynamic perspective, we conclude that initial release styles have limited effects on living fish in unsteady wakes.

\end{abstract}


\end{frontmatter}

\section{Introduction}
\label{Sec_intro}

Fluid flows are complex in natural environments (such as oceans and rivers), usually characterized by unsteady vortices resulting from multiple-body interference (e.g., rocks, corals, fish schooling, etc.). Understanding the swimming/flapping behavior of organisms in complex flows from a fluid mechanics perspective is of importance for exploring natural mysteries and designing more efficient and intelligent bio-inspired robots. Previous studies have primarily categorized the unsteady vortex flows interacted by organisms, such as fish, into two types: avoidance phenomena and schooling phenomena \citep{Wang2019PRE}. For the former one, fish might encounter a fixed bluff body structure (e.g., rocks or corals) while swimming, and it needs to confront a Kármán vortex street shed from upstream. In contrast, in schooling, upstream fish may generate reverse Kármán vortex streets in their wake, requiring followers to adjust their movement patterns and formations accordingly. It has shown that fish can perceive changes in surrounding vortices through sensory organs on either side of their body \citep{Webb1998}. When unsteady vortices are present in the fluid environment, swimming fish can adjust their swimming mode to gain hydrodynamic benefits \citep{Liao2003,Zhu2014PRL}.

For a long period, there has been extensive research on the swimming behavior of organisms in the two typical scenarios mentioned above. \cite{Liao2003,Liao2007} conducted a comprehensive experimental investigation into the influence of vortices generated by a fixed cylinder on trout. They observed the '\textit{Kármán gait}' exhibited by trout across a range of flow velocities. The Kármán gait is a unique swimming pattern adopted by fish behind obstacles, characterized by low-frequency lateral movements and significant lateral oscillations coupled with the vortex street. The lateral motion and body curvature are considerably more pronounced compared to their typical cruising behavior. Importantly, when fish employ the Kármán gait by maneuvering around the vortex cores, they can maintain stable locomotion within complex wakes formed by the upstream cylinder. This results in a reduction in muscle activity and oxygen consumption required for counterflow swimming. In addition to live fish experiments, simplified model experiments, and numerical simulations have also extensively explored the dynamics of a swimming "fish" in the vicinity behind a fixed cylinder. A significant portion of these investigations primarily focuses on the interaction of stationary undulating or flapping plates with cylinders \citep{Zhang2022JFM,Furquan2021JFM,Jia2009POF,Mittal2022JFM}. Only few studies have applied the propelled model to investigate the dynamics and interactions of fish with the bluff body's unsteady wake. Assuming the fish can be modeled by a self-propelled flexible filament or plate, \cite{Park2016POF} and \cite{Wang2019PRE} numerically demonstrated that the propelling filament/plate can spontaneously maintain a stable swimming mode, referred to as the Kármán gait, at a specific downstream location behind fixed cylinders. An interesting phenomenon occurs at the equilibrium position, where the leading edge of the flexible filament/plate consistently crosses around the core of the vortex street, resembling a skier navigating obstacles. Additionally, both studies found that regardless of the position of the self-propelled fin in relation to the upstream stationary cylinder at the beginning, the fin would always choose a few limited positions and hold itself in the action of vortices to save power.

In terms of schooling phenomena, a relatively accepted point is that downstream fish, taking the role of followers, can strategically position themselves within the wakes of leading swimmers to gain hydrodynamic advantages \citep{Weihs1973,Koch2011,Zhu2014PRL,Uddin2015,Park2018JFM,NewboltPNAS2019,Harvey2022SR}. Among these studies, \cite{Zhu2014PRL} simplified fish locomotion as a self-propelled flexible flapping filament and employed numerical simulations to investigate the swimming behavior of two tandem-arranged swimmers. The results indicate that the downstream swimmer often crosses through the vortex core shed by the wake of the upstream swimmer, resulting in reduced energy expenditure during locomotion. Additionally, the gap between the swimmers dynamically changes initially but eventually stabilizes. These findings are consistent with the experimental results of \cite{Ramananarivo2016PRF} and the numerical results of \cite{Park2018JFM}. It is noteworthy here that the self-propelled flexible filament model has become a valuable tool for investigating fish behaviors, allowing swimmers to be dynamically determined through interactions mediated by the surrounding fluid, with both active and passive deformations. Furthermore, swimmers in the self-propelled flexible filament model can couple with variations in the ambient flow, leading to mutual interactions that closely mimic actual biological swimming scenarios \cite{Zhu2014PRL,Peng2018JFM}.

The research mentioned above has clearly demonstrated that fish can couple with the unsteady vortices through specific swimming behaviors to improve swimming efficiency and stability. Note that, in the real nature, the coupling between swimming and the upstream wakes can be sudden or can be developed from the initial state. For example, fish may suddenly encounter obstacles such as rocks or corals or suddenly enter the queue of fish schooling, joining an already fully developed unsteady flow. On the other hand, organisms may also perceive upstream flow in advance and couple with it before the upstream wake fully develops, such as simultaneous initiation for swimming of the schooling fish. All the above examples can be attributed to the (1) different release styles and (2) different initial flow fields faced by organisms. Understanding the effects of these two factors on biological swimming behavior is not yet clear. To the best of our knowledge, to date, only \cite{Wang2019PRE} have examined the swimming behavior of fish behind a fixed cylinder under two different release styles, namely Style 1 and Style 2. The results indicate that the styles significantly affect the final swimming state. In addition, compared to suddenly entering the flow field (Style 2), the Style 1 in \cite{Wang2019PRE} involving advanced perception of flow is more conducive for fish to find energy-saving swimming mode. Furthermore, \cite{Wang2019PRE} attributed the mechanism of these differences to the different styles altering the shedding state of the upstream cylinder and quantitatively analyzed the style effects using the wake phase variations. While \cite{Wang2019PRE}'s study is inspiring, there are still some issues that need further investigation. For example, (1) whether different release styles in schooling phenomena can produce effects similar to those studied previously for the avoidance phenomena? (2) Does the diverse range of release styles present in real-life organisms have varying effects on fish swimming in unsteady wake environments? (3) Importantly, as in avoidance phenomena where fish often couple with fully developed vortex streets rather than encountering wakes evolving from a quiescent flow (as studied in \cite{Wang2019PRE}), whether the mechanisms and phenomena found previously still hold true when using initial flow conditions closer to real flows?

Based on the academic gap mentioned in the last paragraph and inspired by \cite{Wang2019PRE}, we examined the swimming behavior of a downstream fish both in the wake of a fixed upstream cylinder and in the wake of a fish schooling, with a total of seven release styles. Additionally, the above tests were conducted with both quiescent flow fields and fully developed unsteady flows as initial conditions. The results indicate that contrary to previous understandings of this issue, in most conditions, the effects of release styles on swimming behavior in unsteady wakes are limited from a hydrodynamic perspective.

\section{Problem formulation} 
\label{Sec_problem_formulation}

\subsection{Physical problem}
\label{Sec_physical_problem}

\begin{figure}                                                      
	\centering                                                  
	\includegraphics[width=13.5cm]{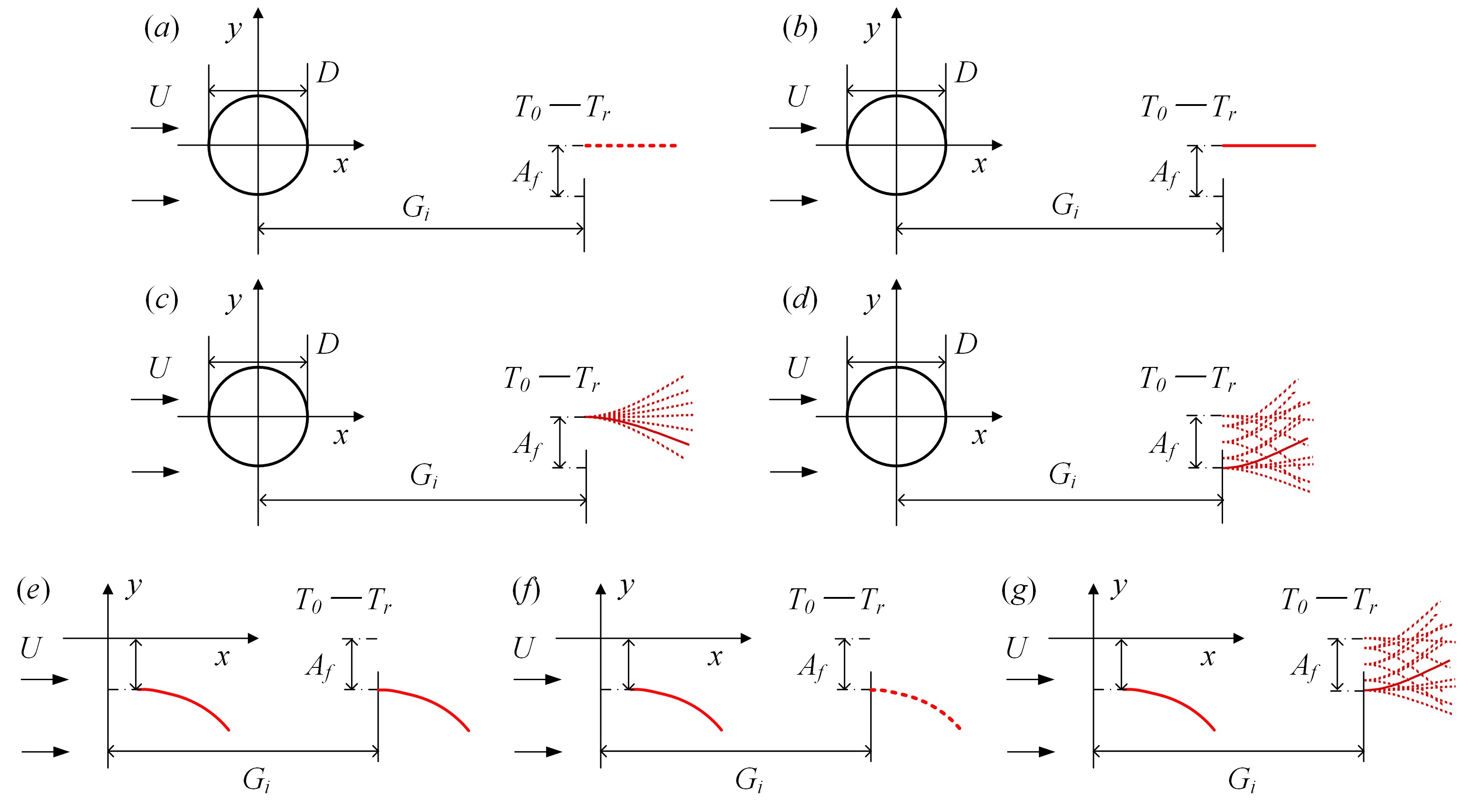}

\caption{A schematic of the problem studied, with (a) a self-propelled fin behind the stationary circular cylinder under flow, and (b) two fins arranged in tandem. SI to SVII represent seven different release styles.}                                     
\label{Fig_schematic}                                               
\end{figure}

The swimming system considered at present is illustrated in Figure \ref{Fig_schematic}. We primarily employ a mechanical model based on the self-propelled two-dimensional massive and inextensible flexible filament/fin of length $L$, to describe the propelled fish. In the present numerical simulations, a vertical oscillatory drive $Y_f(t)$ is imposed at the leading edge of the flexible fin in the direction of the oncoming flow, after the releasing time $T_r$. The resulting flexible deformation, apart from the leading edge, is determined by the fluid-structure interactions (FSI) between the fin and the fluid. This combined active and passive flexible flapping propulsion model is reasonable, as the wake structures generated by the flexible fin have been demonstrated to resemble those of real biological organisms. Furthermore, this approach is widely used to investigate the hydrodynamic characteristics of self-propelled systems \citep{Zhu2014PRL,Peng2018JFM,Park2018JFM}. We here consider seven sets of release styles performed between the initial computational time $T_0$ and $T_r$, namely SI to SVII, and to see their effects on the final dynamics of the fin. As shown in Figure \ref{Fig_schematic}$(a-d)$, SI to SIV are applied for the self-propelled fin in the Kármán street generated by the upstream cylinder. More specifically, SI represents a fin suddenly joining the fully developed wake at $T_r$ and then being propelled; SII represents a fin that keeps stationary between $T_0$ to $T_r$, and then starts propelling after enough interactions with the upstream vortices; SIII represents a fin first passively flaps its body under the fluid-structure interactions during $T_0$ to $T_r$ and then starts to swim at $T_r$; SIV represents a fin self-propelled since $T_0$ but only allowed to move horizontally after $T_r$. Here, SV - SVII are used to estimate the effects of release styles on the dynamics of fish schooling, with two self-propelled fins arranged in tandem, see Figure \ref{Fig_schematic}$(e-g)$. SV represents the two fins that start their locomotion at the same time, while the SVI and SVII are similar to SI and SIV, respectively.   

\subsection{Numerical model}
\label{Sec_numerical model}

The governing equations for the fluid flow in the computational domain are solved by the two-dimensional Navier–Stokes equations, including mass and momentum conservation, as follows:

\begin{align}
	\nabla \cdot \textbf{u} & =0 \label{eq2.1}\\[0.5em]
	\dfrac{\partial \textbf{u}}{\partial t}+(\textbf{u} \cdot \nabla)\textbf{u} &= -\nabla p+\dfrac{1}{Re}\nabla^2\textbf{u}+f, \label{eq2.2}
\end{align}
where $\textbf{u}$ and $p$ represent the non-dimensional velocity and pressure field, scaled based on the incoming velocity $U$ and $\rho_f U^2$ ($\rho_f$ is the density of the fluid), respectively. Here, $Re$ represents the Reynolds number, defined as $Re=UD/\nu$, where $\nu$ is the viscosity. The term $t$ is the non-dimensional time, defined by the real-time $T$ multiply $U/D$. The right-hand term $f$ in Eq. (\ref{eq2.2}) represents the Euler force, nondimensionalized by the $\rho_f U^2/D$. Details of the approach for solving Eqs.(\ref{eq2.1})(\ref{eq2.2}) can be found in \cite{Huang2007JCP}. 

The governing equations for the self-propelled swimmer in the non-dimensional Lagrangian form can be written as  
\begin{equation}
	\rho_s\dfrac{\partial^2 \textbf{X}}{\partial t^2}-\dfrac{\partial}{\partial s}(T_f^* \dfrac{\partial \textbf{X}}{\partial s}) + \dfrac{\partial^2}{\partial s^2}(\gamma \dfrac{\partial^2 \textbf{X}}{\partial s^2}) = -\textit{\textbf{F}},
	\label{eq2.3}
\end{equation}
where $s$ represents the arclength and \textbf{X} (scaled by the reference filament length $L$) is a function of $s$ and $t$, denoting the Lagrangian positions. Here, $\textit{\textbf{F}}$ is the Lagrangian forcing exerted on the fin by the surrounding fluid. The parameters, $T_f^*$ and $\gamma$ represent, respectively, the dimensionless tension force and the bending rigidity, scaled by $\rho_s U^2$ and $\rho_sU^2L^2$. The parameter $\rho_s$ represents the density difference between the fin and the surrounding flow, see \cite{Huang2007JCP} for more details. Due to the inextensible assumption of the filament, we have
\begin{equation}
	\dfrac{\partial \textbf{X}}{\partial s} \cdot \dfrac{\partial \textbf{X}}{\partial s} = 1.
	\label{eq2.5}
\end{equation}

The tension coefficient $T_f^*(s,t)$ in equation (\ref{eq2.3}) can be implicitly solved by the Poisson equation derived from Eqs. (\ref{eq2.3}) and (\ref{eq2.5}), yields
\begin{equation}
	\dfrac{\partial \textbf{X}}{\partial s} \cdot \dfrac{\partial^2}{\partial s^2}\left(T_f^*\dfrac{\partial \textbf{X}}{\partial s}\right) = \dfrac{1}{2}\dfrac{\partial^2}{\partial t^2}\left(\dfrac{\partial \textbf{X}}{\partial s} \cdot \dfrac{\partial \textbf{X}}{\partial s}\right) - \dfrac{\partial^2 \textbf{X}}{\partial t \partial s} \cdot \dfrac{\partial \textbf{X}}{\partial t \partial s}-\dfrac{\partial \textbf{X}}{\partial s} \cdot \dfrac{\partial}{\partial s}(\textbf{\textit{F}}_{b}-\textbf{\textit{F}}).
	\label{eq2.6}
\end{equation}

Following \cite{Zhu2014PRL,Park2018JFM}, a forced harmonic motion $Y_f(t)=A_f\sin(2\pi ft)$ in Y-axis direction is applied on the first Lagrangian point of the fin for the self-propelled model, while all rest Lagrangian points undergo passive motion under the fluid-force. Here, $A_f$ and $f$ are the amplitude and flapping frequency. The swimmer, i.e., the fin, can freely move horizontally and we set its tail to be also free. Based above description, we have following boundary conditions are imposed respectively at the leading and trailing edges, after the the release time $T_r$
\begin{equation}
	Y(t)_{s=0}=A_f\sin(2\pi ft+\theta), \quad \dfrac{\partial \textbf{X}}{\partial s}_{s=0}=(1,0), \quad \dfrac{\partial^3X}{\partial s^3}_{s=0}=0,
	\label{eq2.7}
\end{equation}
\begin{equation}
	T_f^*|_{s=1}=0, \quad \dfrac{\partial^2 \textbf{X}}{\partial s^2}_{s=1}=(0,0), \quad \dfrac{\partial^3 \textbf{X}}{\partial s^3}_{s=1}=(0,0).
	\label{eq2.8}
\end{equation}
Before time $T_r$, the boundary conditions are determined by the release styles accordingly and will be defined later in Eq. (\ref{eq_Bcond_cyl}). 

By combining Eqs. (\ref{eq2.3}) to (\ref{eq2.8}), the deformations and locomotion of the flexible fin in the self-propelled model can be solved. More details for the treatments on the numerical process can be found in \cite{Huang2007JCP}. The present numerical model has been extensively validated previously, with cases of flow over a stationary circular cylinder \cite{Han2022IJHF}, a vortex-induced vibration cylinder \cite{Han2024Arxiv}, a single and tandem propelled fins \cite{Park2018JFM}, a single and tandem flapping filaments \cite{Huang2007JCP,Huang2010JFM,Kim2010JFM}. Based on previous experiences, we set the entire computational domain as $41D$ in length and $32D$ in width, filled with a uniformly distributed Eulerian grid of 2048 in the x-direction. In the span-wise direction, a uniform Eulerian grid is employed within the range [-4, 4] and is stretched otherwise, resulting in a total of 801 grid points. The Lagrangian grid on the circular cylinder consists of 157 points to match the resolution of the Eulerian grid in the fluid component.

\section{Results and discussion} 
\label{Sec_Results}%
Following Figure \ref{Fig_schematic}, the present results will be analyzed and discussed in two parts, of which one is for the cylinder-fin system and another is for the fish schooling system. 

\subsection{Cylinder-fin system} 
For the four release styles mentioned previously in Figure \ref{Fig_schematic}$(a-d)$, we varied the gap between the upstream cylinder and the propelled fin $G_i$ from 4 to 15 under each style. The tested Reynolds number is fixed at $Re=200$, while the fin will start to swim after the release time $T_r$ with an amplitude of $A_f=0.3$. The flapping frequency of the leading edge is the same as the vortex shedding frequency of the upstream cylinder, with $f=0.2$. In each case, the release dimensionless time $T_r$ is 400, which is totally enough for the cylinder to develop its wake and for the fin to interact with the upstream vortices. The bending rigidity $\gamma$ and the mass ratio $\rho_s$ of the fin are 0.2 and 1, respectively. The above applied structural parameters are identical with \cite{Park2016POF}. The flapping amplitude $A_f$ and frequency $f$ of the leading edge are 0.3 and 0.2, respectively. Boundary conditions in this section include the velocity inlet, outflow, and far-field conditions, applied at the inlet, outlet, and upper/lower boundaries, respectively. Between the start time $T_0=0$ and release time $T_r=400$, the boundary conditions of SI and SII are explicit, while those of SIII and SIV are listed as below 
\begin{equation}
\begin{aligned}			
& { \rm SIII}: \textbf{X}_{s=0}=(G_i,0), \quad \dfrac{\partial \textbf{X}}{\partial s}_{s=0}=(1,0) \\
& { \rm SIV}: Y(t)_{s=0}=A_f\sin(2\pi ft+\theta), \quad \dfrac{\partial \textbf{X}}{\partial s}_{s=0}=(1,0), \quad X(t)_{s=0}=G_i,
\label{eq_Bcond_cyl}
\end{aligned}
\end{equation}

\begin{figure}[htb]                                                     
	\centering                                                  
	\includegraphics[]{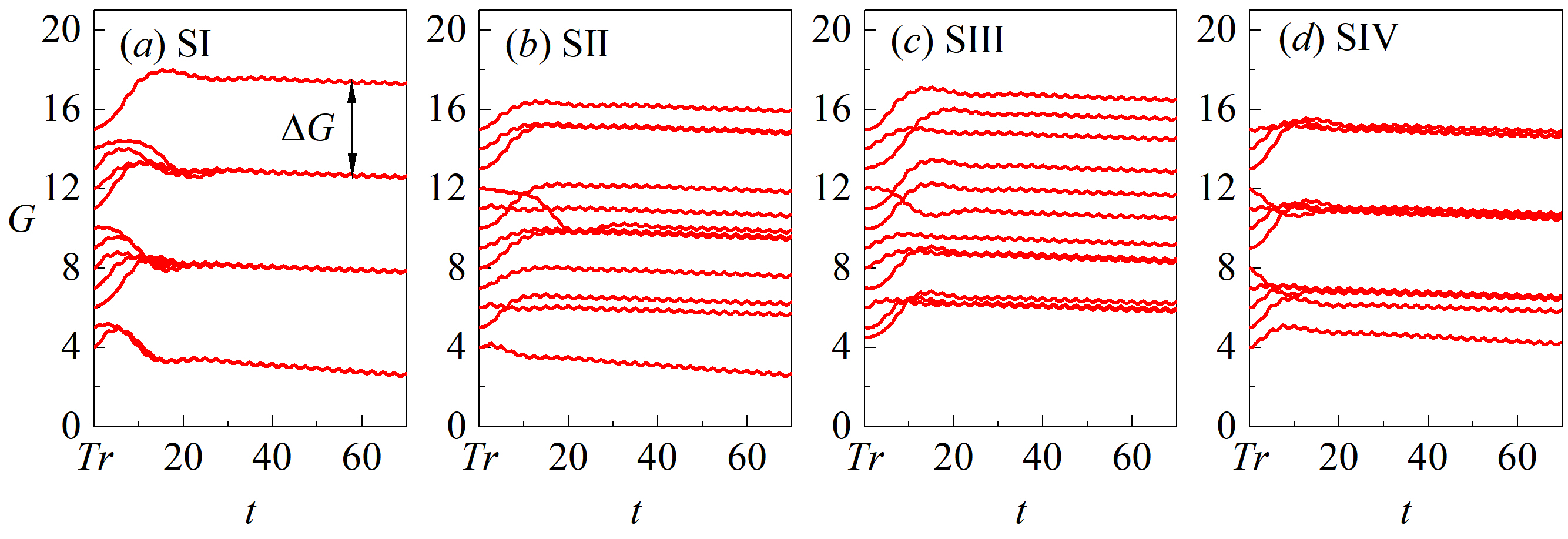}
	
	\caption{Time histories of the dynamic gap $G$ between the upstream cylinder and the fin's leading edge, under different release styles: $(a)$ SI; $(b)$ SII; $(c)$ SIII; $(d)$ SIV.}                                     
	\label{Fig_Gap_u_cyl}                                               
\end{figure} 

Figure \ref{Fig_Gap_u_cyl} shows the time histories of the dynamics gap $G$, under different release styles. In terms of SI, the fin is not able to perceive the surrounding fluid flow, as it suddenly joins the fully developed wake. As a result, regardless of the initial gap, the fin faces the same wake formed by the cylinder. The results indicate that the fin's positions converge to several solutions and the difference among the neighbor stable solutions are basically the same, i.e., $\Delta G\approx4.7$. The value of 4.7 is also the vortex gap among vortices of the wake shed behind a circular cylinder at $Re=200$. The above phenomena are consistent with the findings of \cite{Park2016POF}, indicating the fin is in the Kármán gait state. In addition, \cite{Park2016POF} also proves that the fin holding a stable position behind the cylinder is a sign of saving power compared to the case that a single fin will be pushed downwards by the incoming flow, without a cylinder. Still, under SII, SIII and SIV, the self-propelled fin is also able to stabilize its body in the unsteady wake environment with a certain position and save power, however, the number of converged solutions is more than that in SI. In particular, for SII and SIII, the gap $\Delta G$ between stable positions is unperiodic and becomes much less than 4.7. In addition, we show here for the release of SI, the stable gaps of the fin and the cylinder increase positively (or at least remain unchanged) with the increase in the initial release spacing $G_i$. On the contrary, it can be observed from SII-SIV that the curves may intersect, implying that even if the fin initiates its motion from a position farther away from the cylinder, it may still maintain stability closer to the cylinder. For instance, in SIII, the stable positions for $G_i=11$ and 12 are respectively 12.86 and 10.52. When the fin is released with SII-IV, the smaller spacing $\Delta G$ enables it to acquire more equilibrium positions within the same wake region compared to the scenario with Style I. This, of course, alteration is advantageous as it allows the fins to more easily find equilibrium positions, thereby sustaining stable swimming. In fact, \cite{Stewart2016} mentioned that trout can achieve a Kármán gait at any position within a specific region behind a bluff body to maintain stable swimming, rather than at discrete equilibrium points. 

\begin{figure}[htb]                                                      
	\centering                                                  
	\includegraphics[width=12cm]{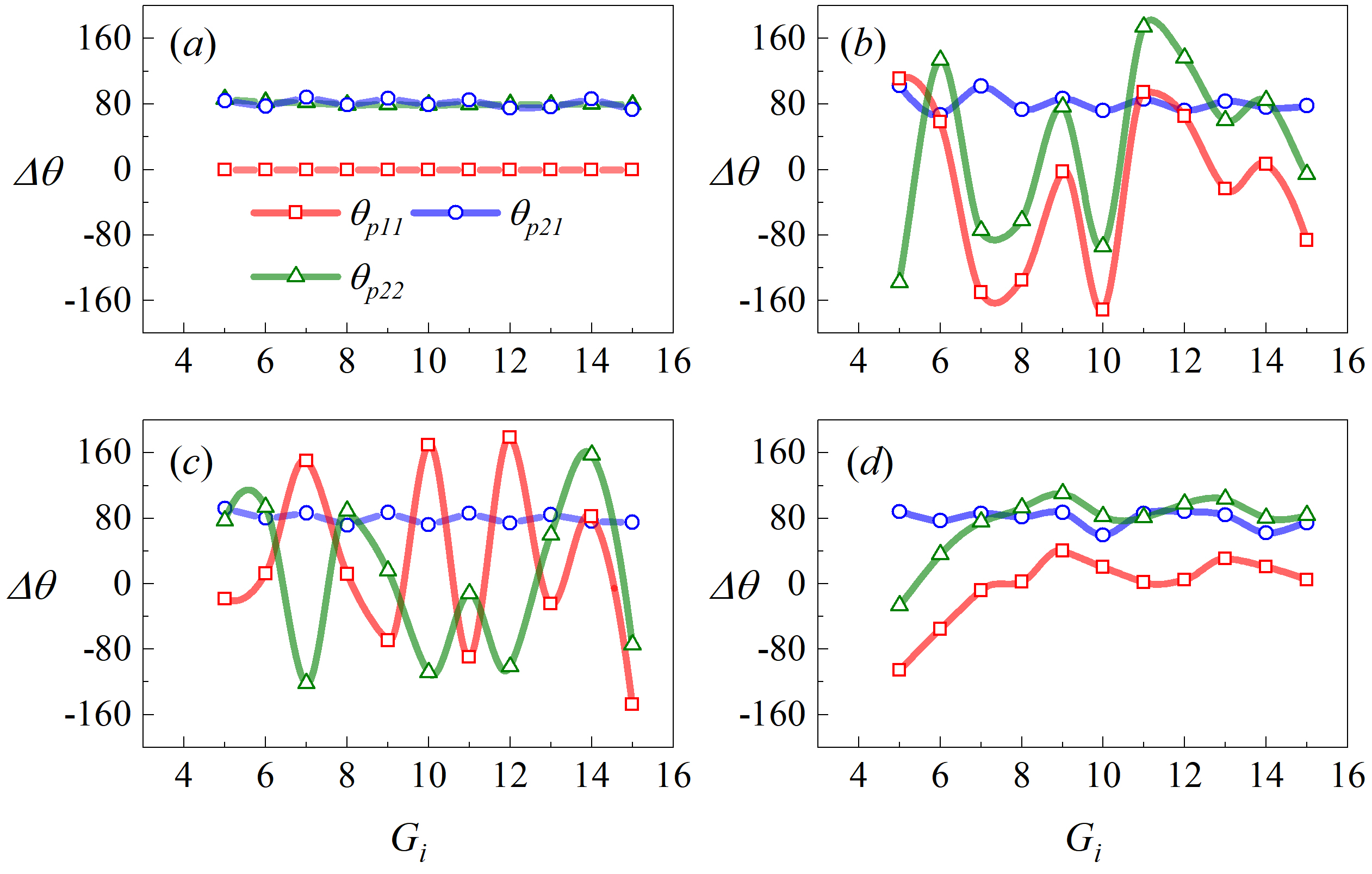}

	\caption{The phase differences of the Y-direction velocity $v$ at points ($4D$, 0) and ($G_i-0.5D$, 0) of the flow field. $(a)-(d)$ Release Style I to IV. The angles $\Delta \theta$ at each $G_i$ represent the relative phase differences compared to the last case of $G_i-1$.}                                     
	\label{Fig_Phase_cyl}                                               
\end{figure}

\cite{Wang2019PRE} tested the response of self-propelled flexible plates released in circular cylinder wake under two different release styles, similar to SI and SIV at present. They assumed that the effects of the initial release styles on the final stable state can be attributed to the phase variation between the vertical velocity of the flow field. Based on the above assumption, \cite{Wang2019PRE} successfully conducted a quantitative analysis through the following equation

\begin{equation}
  \Delta G = \dfrac{2\pi}{\Delta \theta_{p11}/\Delta G_i + 2\pi / \lambda},
	\label{eq_gap}
\end{equation}
where $\lambda$ represents the wake wavelength of vortices shed by a bare circular cylinder. At the present Reynolds number of 200, $\lambda$ equals 4.7. Here, $\Delta \theta$ represents the phase difference used to quantitatively characterize the flow environment. According to \cite{Wang2019PRE}, $\Delta \theta$ can be obtained by monitoring the vertical velocity $v$ of fixed points in the fluid. At present, we consider probe point 1 located at four diameters downstream from the rear of the cylinder along the centerline [$P_1=(4D,0)$], and probe point 2 located at 0.5 diameters upstream from the leading edge of the self-propelled plate at the release time [$P_2=(G_i-0.5D,0)$]. Clearly, the point 2 will vary with different initial release gaps $G_i$. Based on the velocities at these two points, three phase angles $\Delta \theta$ can be defined: (1) the phase difference $\theta_{p11}$ between probe points 1 and 1, and (2) the phase $\theta_{p22}$ between points 2 and 2, under different initial release gaps $G_i$, i.e., tested cases; (3) the phase difference $\theta_{p21}$ between probe point 2 and point 1, under the same $G_i$. Figure \ref{Fig_Phase_cyl} presents the variation of those above three phase differences with the initial spacing $G_i$ under different release styles. Note that, we take the data at $G_i=4$ as the reference value, where the phase difference $\Delta \theta$ is considered as 0. In addition, the angles at other $G_i$ represent the relative phase differences compared to the last gap case $G_i-1$. It can be observed that the phase $\theta_{p11}$ is essentially affected by the other two phase differences $\theta_{p22}$ and $\theta_{p21}$, and theoretically, $\theta_{p21}=\theta_{p22}-\theta_{p11}$. For the SI release style, the phase $\theta_{p22}$ remains 0, indicating that the fluid environment encountered by the self-propelled fin at each $G_i$ remains exactly the same. Therefore, $\theta_{p22}$ equals $\theta_{p21}$, indicating that the $G_i$ is the only factor affecting the phase of this system for all the tested cases. One can find that, for each increment of $G_i$ by 1, $\theta_{p21}$ and  $\theta_{p11}$ increase by approximately $80^\circ$ degrees (or $4\pi/9$) and 0, respectively. Based on the uniform increase in the above $\theta_{p21}$, one cycle of $2\pi$ corresponds to an increment of $\Delta G$ of 4.5, indicating that with the $G_i$ varying from any value $\delta$ to $\delta+4.5$, the fin's leading edge experiences identical fluid flow and it is expected to behave similarly. Also, the value of 4.5 agrees well with the computed $\Delta G$ in Figure \ref{Fig_Gap_u_cyl}($a$) and also the analytical results obtained from Eq. (\ref{eq_gap}). 

For the release style SIV, it can be observed that before $G_i$ falls below 7, there is considerable fluctuation in the phase difference, whereas with the increase in $G_i$, the phase difference tends to stabilize, with an average value around 89.7 degrees ($0.5\pi$). Assuming that the minor fluctuations when $G_i$ is greater than 7 can be disregarded, then within one cycle, it corresponds to a gap variation of 4, which is consistent with the results obtained from Figure \ref{Fig_Gap_u_cyl}($d$)(4.1) and Eq. \ref{eq_gap}. These results support the view by \cite{Wang2019PRE} that the release style effects are related to the phase difference perceived by the fish in the flow field. However, it is worth noting that under release styles SII and SIV, the above conclusion no longer holds. For instance, seeing cases of $G_i=11$ and 12 from Figure \ref{Fig_Gap_u_cyl}(b), a phase difference of $65.2^\circ$ is obtained, yielding a gap between the two stable solutions of $2\pi/(0.365\pi+2\pi/4.7) = 2.54$ based on Eq. \ref{eq_gap}. However, numerical results indicate a value of $\Delta G=-0.79$. Similarly, for SIII, see the two cases $G=11$ and $G_i=12$ from Figure \ref{Fig_Gap_u_cyl}($c$) yields a result of 1.41 based on Eq. \ref{eq_gap}, which is inconsistent with the numerical simulation result of -2.33. Observing the characteristics of SII and SIII in Figure \ref{Fig_Phase_cyl}($b$) and ($c$), it is evident that the phase differences $\theta_{11}$ and $\theta_{22}$ vary significantly, indicating an influence of the fin on the upstream cylinder's flow field under these release styles. Moreover, based on the aforementioned analysis, we also consider that the effects of release SII and SIII on stable positions cannot be simply explained by the single phase difference $\theta_{11}$ in the flow field. Combining with Figure \ref{Fig_Gap_u_cyl}, it can also be observed that there is no good convergence for the stable solutions under SII and SIII, which is inherently related to the fluctuation of the phase difference in Figure \ref{Fig_Phase_cyl}.

\begin{figure}[htb]                                                      
	\centering                                                  
	\includegraphics[width=12cm]{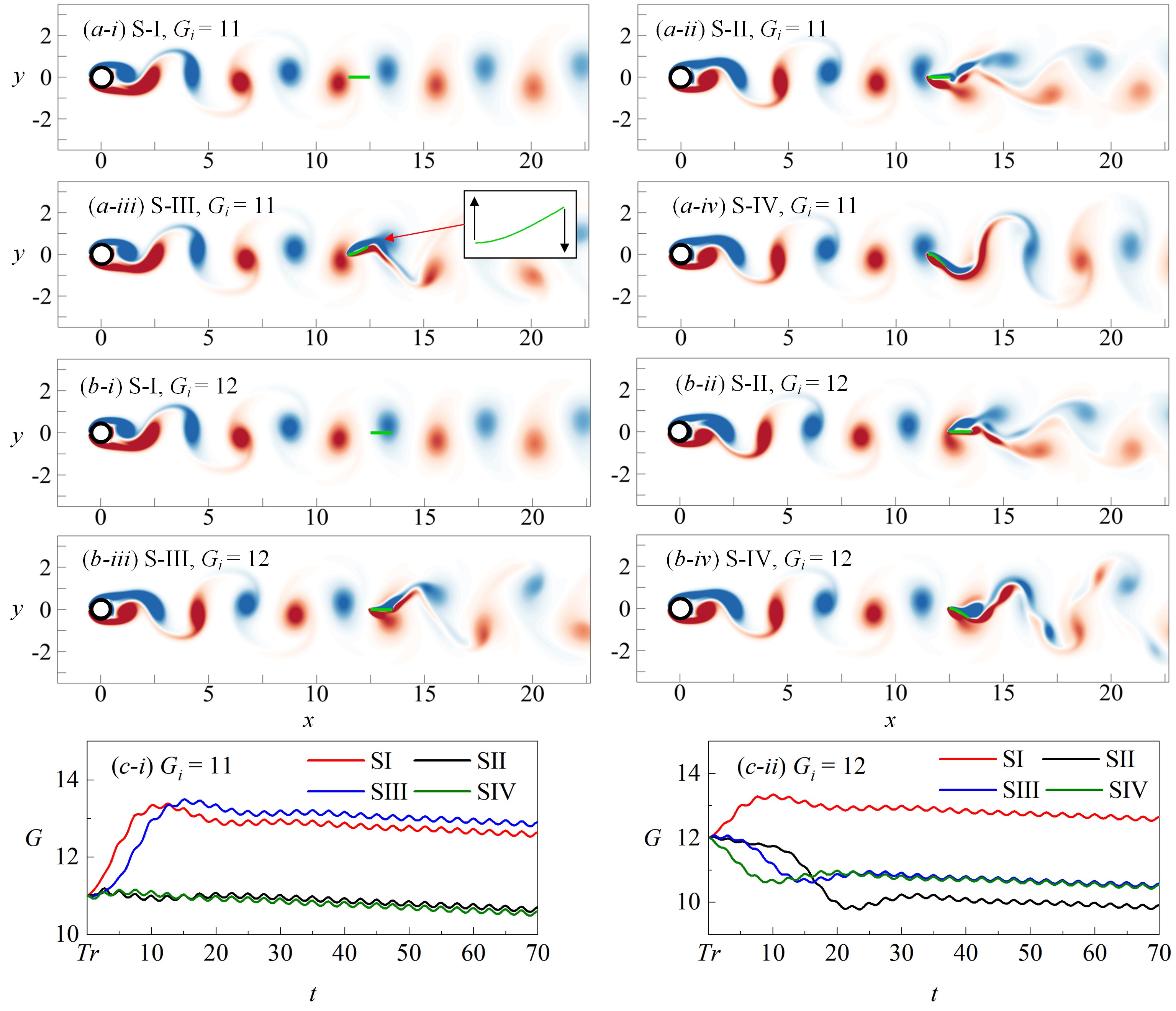}

	\caption{($a$) and ($b$) Vorticity contours at the release time $T_r$, under different release styles. ($c$) Time histories of the dynamic gap $G$ between the upstream cylinder and the fin's leading edge, under different release styles, at $G_i=11$ and 12 (data is taken from Figure \ref{Fig_Gap_u_cyl}). }                                     
	\label{Figure_vorz_cyl}                                               
\end{figure}

For the four release styles studied here, we selected two typical gaps, $G_i=11$ and $G_i=12$, to see the interaction between downstream fin and the upstream vortices at release time $T_r$. As shown in Figure \ref{Fig_Gap_u_cyl}($a$), for $G_i=11$ and $G_i=12$, the fin behind the cylinder tends to converge to the same stable solution in SI, while overlapping occurs in SII to SIV. Figure \ref{Figure_vorz_cyl} illustrates the instantaneous vorticity fields at time $T_r$ for different release styles and $G_i$. Since the fin in SI is suddenly introduced into the cylinder's wake, prior to $T_r$, there is no coupling between the fin and the fluid. This implies that the flow in Figure \ref{Figure_vorz_cyl}($a-i$) should be consistent with the flow around a single cylinder at the same $Re$ and can thus serve as reference data for comparison. From Figures \ref{Figure_vorz_cyl}($a-ii$) and ($a-iv$), it can be noted that due to the coupling between the fin and the flow field under SII and SIV, not only the flow states near and behind the fin are altered, but also the shedding of vortices from the upstream cylinder is significantly affected. Therefore, for $G_i=11$, the swimming behavior of SII and SIV should differ greatly from SI, which is consistent with the data in Figure \ref{Figure_vorz_cyl}($c-i$). In terms of $G_i=12$ shown in Figure \ref{Figure_vorz_cyl}($b$), compared to SI, all other release styles SII-SIV significantly alter the shedding state of upstream vortices, indicating an influence on the cylinder shedding. One might find that, comparing Figures \ref{Figure_vorz_cyl}($a-i$)($a-iii$), ($a-ii$)($a-iv$), and ($b-ii$)-($b-iv$), at time $T_r$, the flow environment faced by the fin is similar, leading to similar trajectories for these three sets of cases in Figure \ref{Figure_vorz_cyl}($c$). It is important to note that for SI and SIII modes under $G_i=11$, as well as SIII and SII/SIV at $G_i=12$, although the flow environment faced by the fin is consistent, there are still some differences in the final swimming states. We believe that the influence of different release styles on the results cannot be solely interpreted through flow phase variations. At release time $T_r$, the initial postures of the fin also differ. For instance, in Figure \ref{Figure_vorz_cyl}($a-iii$)($b-iii$), the tail of the fin is raised and tends to flap in the $-y$ direction, while the head is about to initiate the trajectory of imposed motion $Y(t)$ and shows a trend towards motion in the $+y$ direction. It can be considered that the phase difference in the different strokes between the leading and trailing edges will also affect the swimming dynamics of the self-propelled fin.

\begin{figure}[htb]                                                      
	\centering                                                  
	\includegraphics[]{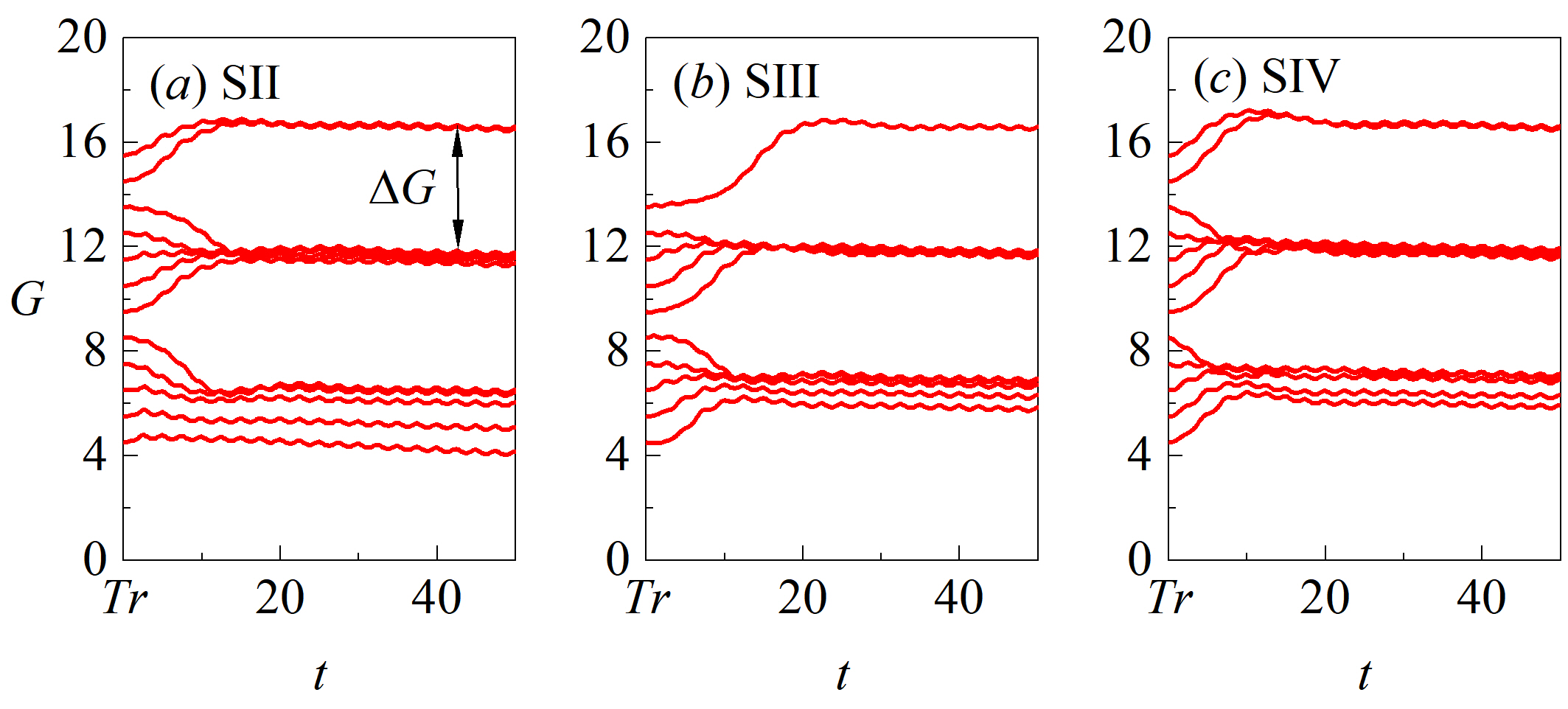}

	\caption{Time histories of the dynamic gap $G$ between the upstream cylinder and the fin's leading edge, under different release styles: $(a)$ SII; $(b)$ SIII; $(c)$ SIV. Note that, the initial data for the simulation is obtained from the fully developed Kármán street of a bare cylinder.}                                     
	\label{Fig_full_Gap_cyl}                                               
\end{figure}

Note that, at the initial time $T_0$ of the simulations mentioned above, the entire flow field of the system is initialized, with the overall pressure set to 0 and the velocity at each grid point set to (1,0). This implies that the downstream fin has coupled with the cylinder and participated in the flow development from beginning to end, even before the wake has fully developed. However, such simulation settings lack physical realism. In the natural environment, fish are often coupled with fully developed unsteady vortex environments and rarely encounter wakes evolving from a quiescent flow. To address this, applying the same parameters as those in Figure \ref{Fig_Gap_u_cyl}, we recalculated the positions of the fin's leading edge under different release styles and initial gap $G_i$, as shown in Figure \ref{Fig_full_Gap_cyl}. Here, we first computed the shedding of vortices around a bare cylinder and used the fully developed results as the initial value at time $T_0$ for subsequent calculations in Figure \ref{Fig_full_Gap_cyl}. This computational approach better reflects the actual conditions encountered by real organisms. Interestingly, when $G_i$ exceeds 6, compared to Figure \ref{Fig_Gap_u_cyl}, we observe that, regardless of the release styles in which the downstream fin applies, it does not affect the final stable outcome. We show that when using a fully developed wake as the initial condition, the fin's position does not exhibit the characteristic of multiple irregular discrete solutions as shown in Figure \ref{Fig_Gap_u_cyl}. Furthermore, by examining the discrete gap in Figure \ref{Fig_full_Gap_cyl}($a$)-($c$), we find that $\Delta G$ all match approximately the value of 4.7 under SI and also the wavelength of Kármán street. This proves again that different release styles seem to not have strong effects on the final swimming behavior, for a more real flow condition. When $G_i$ is less than 6, we see the fin is still influenced to a certain extent by the release styles. This is because, the downstream fin is near the region of vortex separation from the upstream cylinder, allowing it to affect and alter the flow.

\subsection{Fish schooling system} 

The followers in fish schooling must deal with the unsteady vortices generated by the upstream leaders' body movements. In nature, followers typically do not remain constantly within the queue; they may take breaks for rest and adjustments at certain times, then rejoin the queue later. These phenomena correspond to different release styles. This section will examine three typical styles depicted in Figure \ref{Fig_schematic}($d$)-($f$) to address whether these styles can impact the hydrodynamic behavior of fish schooling.

\begin{figure}[htb]                                                     
	\centering                                                  
	\includegraphics[]{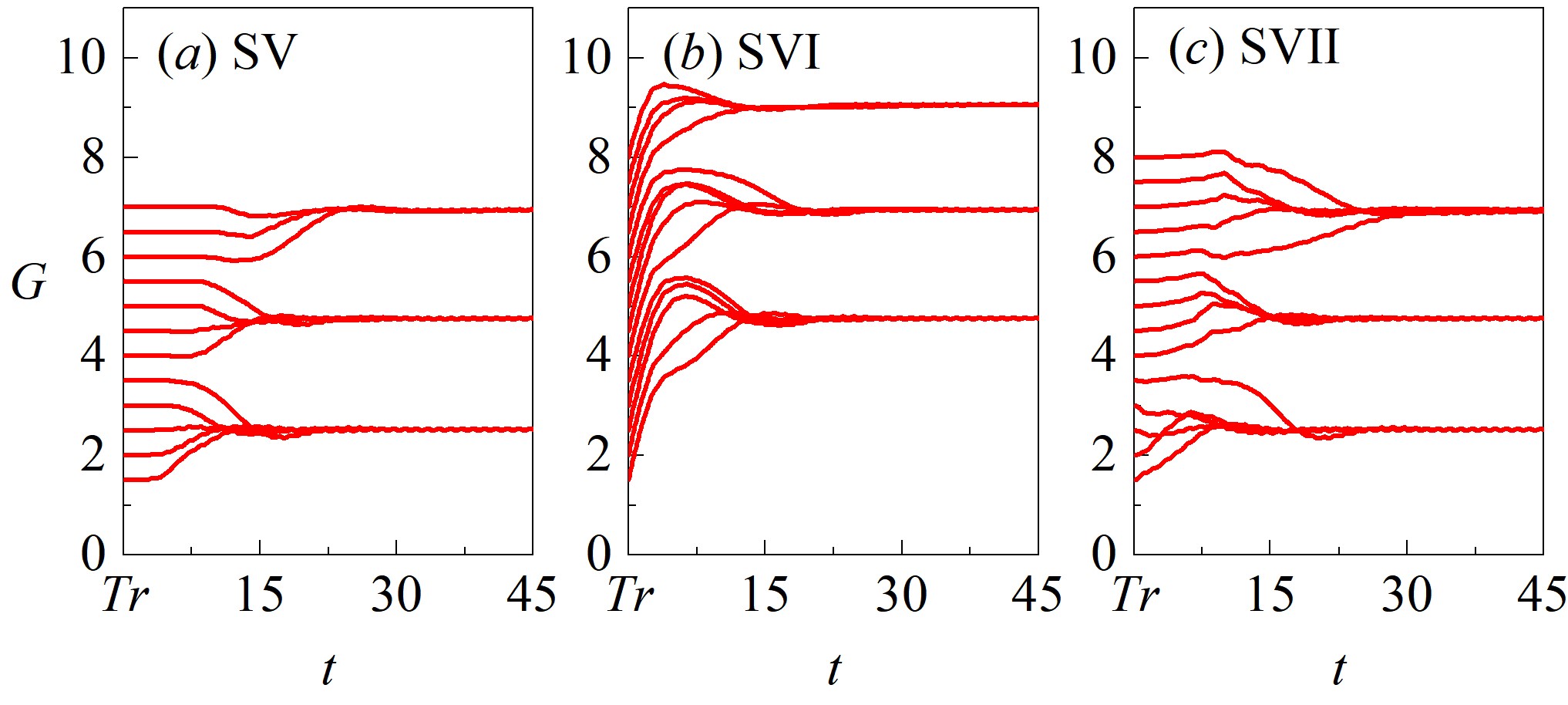}
	
	\caption{Time histories of the dynamic gap $G$ between the upstream and downstream fins in fish schooling, under different initial release styles: $(a)$ SV; $(b)$ SVI; $(c)$ SVII.}                                     
	\label{Fig_Gap_t_fins}                                               
\end{figure}

The simulation settings applied here are as follows: To construct a quiescent flow environment suitable for schooling behavior, we set the inlet velocity to 0 and apply Neumann boundary conditions both at the upper and lower boundaries. In the self-propelled model, the fin's stiffness and mass ratio are both set to 1, the flapping frequency is 0.4, and the flapping amplitude is 0.2. These conditions are similar to those in \cite{Park2018JFM}. For the SV style, no action is taken from time $T_0$ to $T_r$, and the two fins start propelling at $T_r$. For the SVI, the upstream leader starts its locomotion at $T_0$, while the downstream follower suddenly joins the wake at $T_r$. In the SVII style, the upstream fin follows SVI, while the downstream follower flaps between $T_0$ and $T_r$ to couple with the upstream vortices and is only allowed to free to move along X-direction at $T_r$. Since we initiate the simulation at SV from $T_r$ and at SVI and SV cases from $T_0$, the simulation strategy for the fish schooling system simultaneously incorporates the mechanisms related to the initial flow differences as mentioned in Figures \ref{Fig_Gap_u_cyl} and \ref{Fig_full_Gap_cyl}. Figure \ref{Fig_Gap_t_fins} shows the variation of the leader-follower gap $G$ over time for different release styles during fish schooling. As in Figure \ref{Fig_Gap_t_fins}($a$), after a period of adjustment, the spacing $G$ between the two fins in the schooling pattern stabilizes and converges to a few regular solutions. The difference between adjacent convergent solutions, calculated as $\Delta G$, is all around 2.2. These results are consistent with those found in \cite{Zhu2014PRL} and \cite{Park2018JFM}. Previous studies focused on the SI style, but when we change the release styles, we find that the qualitative conclusions remain unchanged. Additionally, the differences between adjacent solutions in Figure \ref{Fig_Gap_t_fins}($b$) and \ref{Fig_Gap_t_fins}($c$) also maintain quantitative consistency with Figure \ref{Fig_Gap_t_fins} ($a$). Comparing Figure \ref{Fig_Gap_t_fins}($a$) and ($b$), it can be observed that for the same initial gap $G_i$, the results of SII may converge to a larger gap, such as $G_i=2.5$ corresponding to $G=2.54$ in SI and to $G=4.77$ in SII. This phenomenon occurs because in the SII mode, when the downstream fin starts its locomotion, the upstream leader already has a swimming velocity $U_c$, which enlarges the gap between the stable states. In conclusion, we believe that in a fish schooling system, regardless of the initiation release styles and initial flow field, it has limited effects on the final swimming behavior of the downstream follower.

\begin{figure}[htb]                                                     
	\centering                                                  
	\includegraphics[width=12cm]{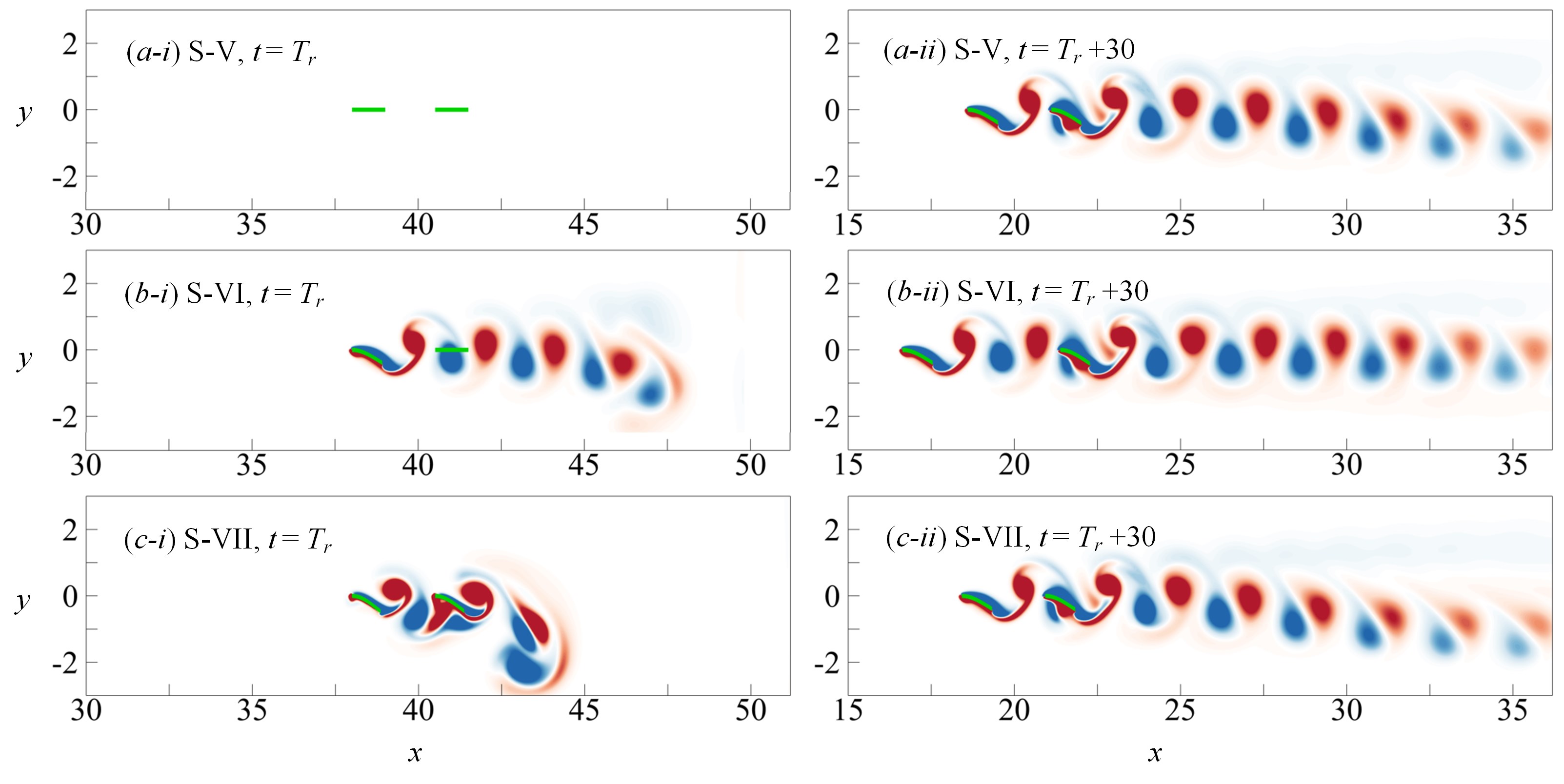}
	
	\caption{Vorticity contours ($a-i$) to ($c-i$) at the release time $T_r$ and ($a-ii$) to ($c-ii$) at the time $T_r+30$, under different release styles SV-SVII, with an initial gap $G_i$ of 2.5.}                                     
	\label{Fig_vor_fins}                                               
\end{figure}

Figure \ref{Fig_vor_fins} shows the instantaneous vorticity fields of the schooling system at time $T_r$ and $T_r + 30$ for different initiation styles with an initial spacing of $G_i=2.5$. Comparing Figures \ref{Fig_vor_fins}($a$) and ($c$), it can be observed that the shedding of vortices during the stable phase is consistent between the two sub-figures. Due to the velocity difference between the upstream and downstream fins when initiated with the SII mode at $Y_r$, the stable spacing differs from that of SI and SIII. Consequently, as shown in Figure \ref{Fig_vor_fins}($b$), the corresponding shedding states also vary. However, it is notable that regardless of the initiation styles, the follower fish are seen crossing the negative vortex cores and preparing to traverse the next alternating positive vortex core. This crossing of vortex cores in swimming behavior is inherently advantageous in terms of hydrodynamic benefits \citep{Zhu2014PRL}. Thus, regardless of the initiation styles, fish tend to adopt energy-saving swimming strategies.

\begin{figure}[htb]                                                     
	\centering                                                  
	\includegraphics[]{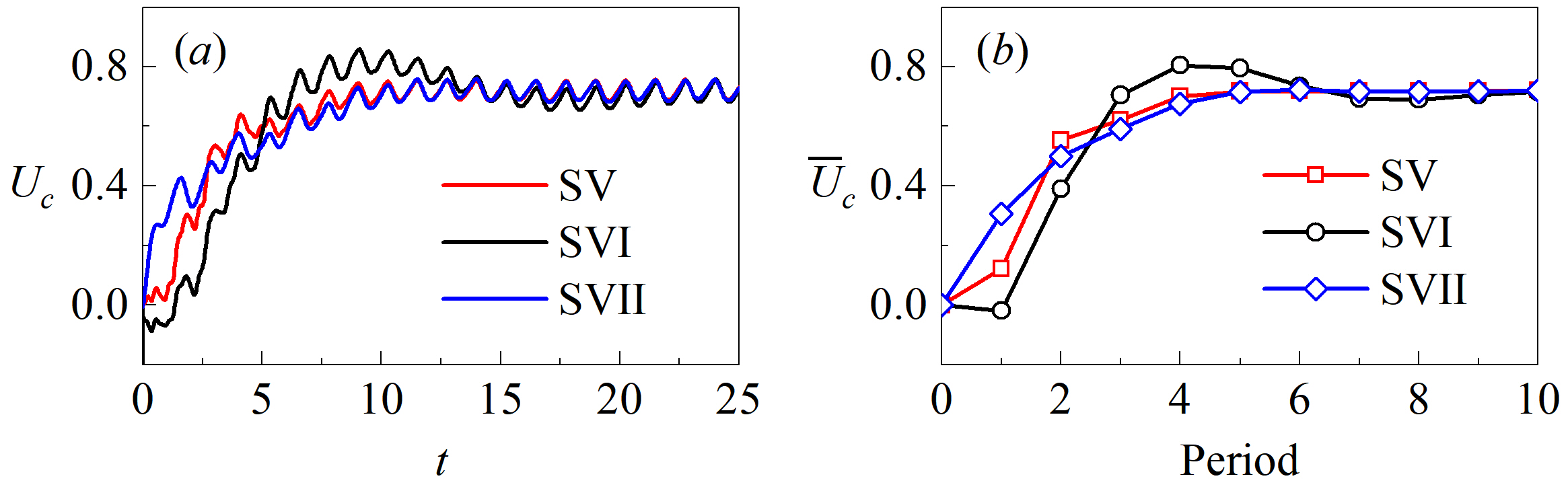}
	
	\caption{($a$) The centre velocity $U_c$ of the follower fin in schooling, with respect to the time under different release styles with $G_i=2.5$. ($b$) The averaged centre velocity $U_c$ at each flapping period $1/f$.}                                     
	\label{Fig_Uc_fins}                                               
\end{figure}

Figure \ref{Fig_Uc_fins} presents the time histories of the swimming velocity $U_c$ of the follower fin in schooling, under different initial gaps $G_i$ and initiation styles. Here, $U_c$ is defined as the velocity at the center Lagrangian point on the fin in the direction of flow. The results indicate that although initiation styles and initial flow fields do not significantly affect the qualitative aspects of the final swimming state, they still influence the swimming velocity during the adjustment phase. However, integrating the swimming velocity over time yields similar results, suggesting that the differences in the velocities also do not significantly impact the final swimming distance. These findings suggest that unsteady fluid vortices play a modulating role in schooling systems.

\section{Conclusions} 
Living fish may suddenly encounter obstacles, join the queue of a fish school, or detect upstream flow in advance, leading to interactions with environmental flows that can either occur abruptly or develop gradually from an initial state. This study aims to address, from a hydrodynamic perspective, the extent to which initial conditions can influence the locomotion characteristics of fish.

We numerically investigate the effects of different initial gaps, release styles, and flow fields (quiescent flow fields and fully developed unsteady flows) on fish swimming in unsteady wakes generated by both upstream bluff bodies (avoidance phenomena) and leading flapping fish (schooling phenomena). The study employs a self-propelled flexible fin model for the swimming fish, direct simulation techniques for fluid flow, and the immersed boundary method for fluid-structure interactions. The results indicate that fish in schooling phenomena are more resilient to initial conditions compared to those in avoidance phenomena. Notably, while different release styles can alter locomotion in avoidance phenomena, the effects of these styles become negligible when simulations start from a fully developed flow field, which more accurately reflects natural conditions. In summary, contrary to recent studies, our findings demonstrate that with more realistic settings, initial release styles have limited effects on the hydrodynamic behavior of a self-propelled fin in unsteady wakes generated by both avoidance and schooling phenomena.

\section*{Acknowledgement}
\label{sec:acknowledgements}
This work was supported by the National Natural Science Foundation of China (No. 12302317, 92252204, and 12272206), and China Postdoctoral Foundation (No. 2023M741883).

\bibliographystyle{jfm2}
\biboptions{authoryear}
\bibliography{Finalrefs.bib}
	
\end{document}